\documentclass[aps,pra,twocolumn,floats,amsmath,amssymb,superscriptaddress]{revtex4-2}
\usepackage[utf8]{inputenc}
\usepackage{graphicx}
\usepackage{epsfig}
\usepackage{amsfonts}
\usepackage{amsmath}
\usepackage{natbib}
\usepackage{multirow}
\usepackage{bm}
\usepackage{epstopdf}
\usepackage{ulem}
\usepackage{color}

\definecolor{/green}{rgb}{0.0, 0.72, 0.92}

\newcommand{\ie}{{\it i.e.}, }

\begin{document}
\title{How ferroelectric BaTiO$_3$ can tune a two-dimensional electron gas at the interface of LaInO$_{3}$ and BaSnO$_{3}$: a first-principles study}
\author{Le Fang}
\affiliation{Materials Genome Institute, International Center for Quantum and Molecular Structures, Physics Department, Shanghai University, 200444 Shanghai, China}
\affiliation{Institut f\"{u}r Physik and IRIS Adlershof, Humboldt-Universit\"{a}t zu Berlin, 12489 Berlin, Germany}
\author{Wahib Aggoune}
\affiliation{Institut f\"{u}r Physik and IRIS Adlershof, Humboldt-Universit\"{a}t zu Berlin, 12489 Berlin, Germany}
\author{Wei Ren}
\email{renwei@shu.edu.cn}
\affiliation{Materials Genome Institute, International Center for Quantum and Molecular Structures, Physics Department, Shanghai University, 200444 Shanghai, China}
\affiliation{Shanghai Key Laboratory of High Temperature Superconductors and State Key Laboratory of Advanced Special Steel, Shanghai University, Shanghai 200444, China}
\author{Claudia Draxl}
\email{claudia.draxl@physik.hu-berlin.de} 
\affiliation{Institut f\"{u}r Physik and IRIS Adlershof, Humboldt-Universit\"{a}t zu Berlin, 12489 Berlin, Germany}
\affiliation{European Theoretical Spectroscopy Facility (ETSF)}

\begin{abstract}
The emerging interest in two-dimensional electron gases (2DEGs), formed at interfaces between two insulating oxide perovskites poses crucial fundamental question in view of future electronic devices. In the framework of density-functional theory, we investigate the possibility to control the characteristics of the 2DEG formed at the LaInO$_3$/BaSnO$_3$ interface by including a ferroelectric BaTiO$_{3}$ layer. To do so, we examine how the orientation of the ferroelectric polarization impacts density and confinement of the 2DEG. We find that aligning the ferroelectric polarization toward (outward) the LaInO$_3$/BaSnO$_3$ interface leads to an accumulation (depletion) of the interfacial 2DEG. Varying its magnitude, we find a linear effect on the 2DEG charge density that is confined within the BaSnO$_3$ side. Analysis of the optimized geometries revels that inclusion of the BaTiO$_{3}$ block makes structural distortions at the LaInO$_3$/BaSnO$_3$ less pronounced, which, in turn, enhances the 2DEG density. Thicker ferroelectric layers allow for reaching higher polarization magnitudes. We discuss the mechanisms behind all these findings and rationalize how the characteristics of both 2DEG and 2D hole gases can be controlled in the considered heterostructures. Overall, our results can be generalized to other combinations of ferroelectric, polar, and nonpolar materials.
\end{abstract}
 
\date{\today}
\maketitle
\section{INTRODUCTION}
Interfaces between oxide perovskites exhibit novel physical phenomena, such as superconductivity~\cite{Reyren+07sc}, magnetoelectric coupling~\cite{yin+2020Nano}, spin polarization~\cite{fred+2015PRB,pent+2009PRL}, or the formation of two-dimensional electron gases (2DEGs). The latter, discovered first at the interface between the two insulating perovskites LaAlO$_3$ and SrTiO$_3$~\cite{Ohtomo+04n}, refers to the accumulation of electronic charge tightly confined in a small region at or across the interface. As this charge can move freely inside the interfacial planes, 2DEGs have attracted tremendous interest in view of their exploitation in the next generation of electronic devices. In this specific example, LaAlO$_3$ is a polar material that consists of alternating charged (LaO)$^+$ and (AlO$_2$)$^-$ planes, while the nonpolar material SrTiO$_3$ is made of charge-neutral (TiO$_2$)$^0$ and (SrO)$^0$ layers. The formation of the 2DEG is a direct consequence of electronic reconstruction, \ie through charge transfer from the LaAlO$_3$ to the SrTiO$_3$ side, which occurs to compensate the polar discontinuity at the interface~\cite{Nakagawa+06nm}. This charge is hosted by Ti-\textit{d} states. For a complete compensation of the formal polarization (termed $P_{\mathrm{f}}$), induced within LaAlO$_3$, it can reach a free charge-carrier density of $\sim$3.3$\times$10$^{14}$ cm$^{-2}$~\cite{Huijben+09am}. This corresponds to 0.5 electrons (e) per unit-cell area.

Spurred by practical applications, many efforts have been made to control and functionalize such interfaces using external stimuli. This includes light illumination~\cite{irvin+2010NP}, strain~\cite{bark+2011PNAS}, or external electric ﬁelds~\cite{Thiel+06sc,cen+2008NM,bi+2010APL,xie+2011AM,wu+2013PRX}. In parallel, theoretical predictions of 2DEGs at interfaces between SrTiO$_{3}$ and ferrorelectric materials have been reported~\cite{Niranjan+09prl,Zhang+11apl}. It was shown that switching the polarization direction, offered by the ferroelectric material, allows for controlling the interfacial charge. Tuning the 2DEG formed at the LaAlO$_{3}$/SrTiO$_{3}$ interface was successfully demonstrated by addition of a ferroelectric functional layer of Pb(Zr,Ti)O$_3$ layer~\cite{kim+2013AM}. More specifically, it was shown that the 2DEG can be reversibly turned \textit{on} and \textit{off} in a non-volatile manner with a large \textit{on/off} ratio ($\textgreater$1000). This exciting finding opened up new perspectives for understanding and exploiting such oxide combinations toward achieving high-speed ferroelectric gate field-effect transistors (FeFET) for non-volatile devices~\cite{kim+2013AM,tra+2013AM,kim+2015Sr}. More fascinating experimental studies on SrTiO$_3$-based heterostructures have been reported, using ferroelectric materials, such as Pb(Zr,Ti)O$_3$~\cite{tra+2013AM,wang+2018ACS,uzun+2020JAP} or BiFeO$_3$~\cite{gao+2018AMI}, to tune its interfacial 2DEG. Moreover, a theoretical study~\cite{gao+2016PE} showed that ferroelectricity in BaTiO$_{3}$ can be used to control the 2DEG at the LaAlO$_3$/SrTiO$_3$ [001] interface through altering the intrinsic electric field of LaAlO$_3$. 
\begin{figure*}[htb]
 \begin{center}
\includegraphics[width=.9\textwidth]{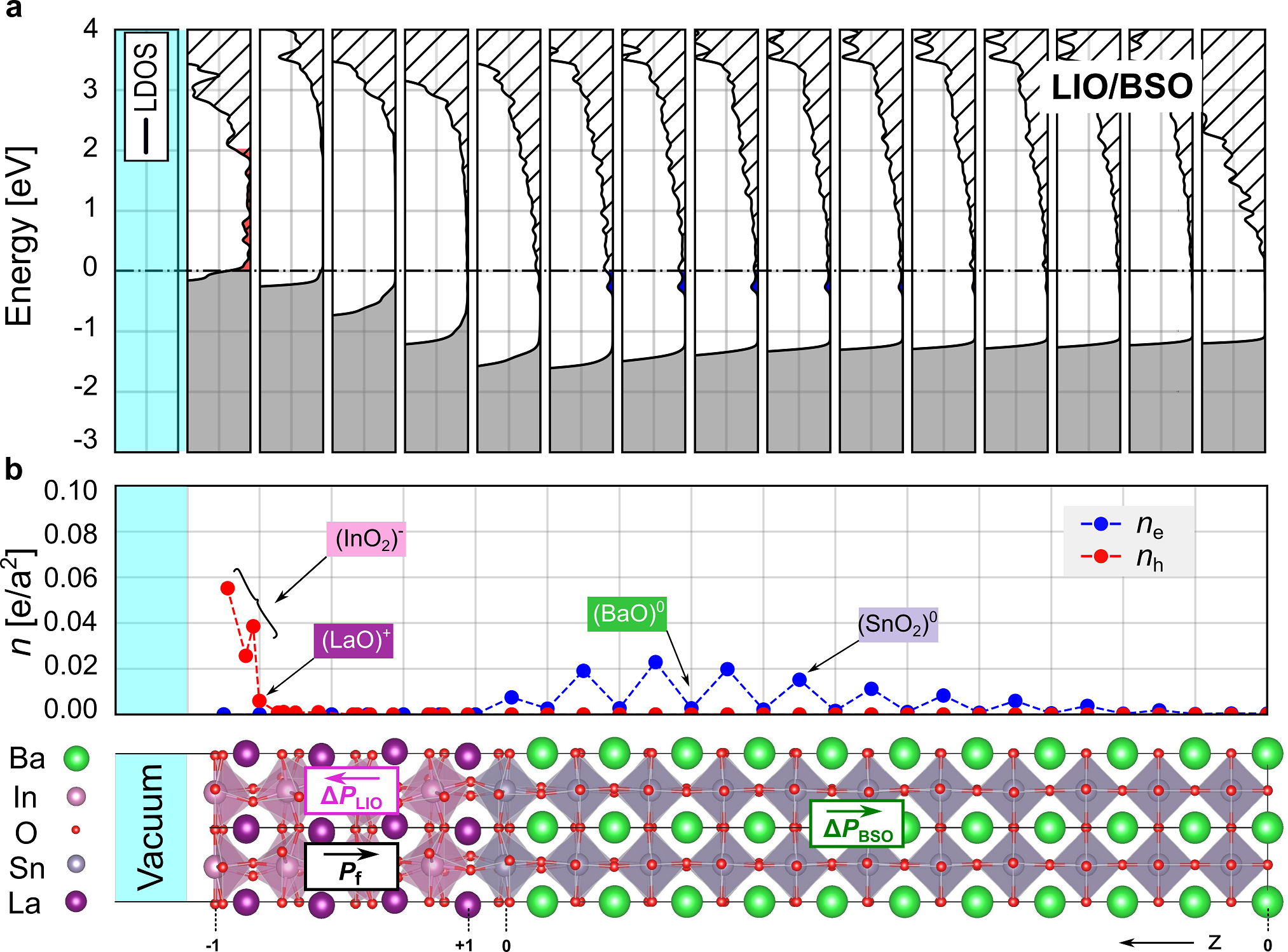}%
\caption{Bottom: Stoichiometric non-periodic LIO/BSO interface formed by 11 BaSnO$_{3}$ unit cells and four LaInO$_{3}$ pseudcubic unit cells in [001] direction (\textit{z}), that is shared between all panels. $P_{\mathrm{f}}$ is the formal polarization oriented from the (InO$_{2}$)$^{-1}$ plane towards the (LaO)$^{+1}$ plane at the interface. $\Delta P_{\mathrm{LIO}}$ (magenta) and $\Delta P_{\mathrm{BSO}}$ (green) are the polarizations due to structural distortions within LaInO$_{3}$ and BaSnO$_{3}$, respectively. (a) Local density of states per unit cell (LDOS) with the Fermi level being set to zero. The shaded gray area indicates the occupied valence states. The depleted valence band region (holes) and the occupied conduction band region (electrons) resulting from electronic reconstruction, are highlighted by red and blue color, respectively. (b) Distribution of the electron (hole) charge densities obtained by integrating the LDOS indicated by blue (red) color in panel (a).
}
\label{Fig-1}
 \end{center}
\end{figure*}

Understanding the microscopic mechanisms affecting the 2DEG in such ferroelectric/polar/nonpolar heterostructures, requires decoupling different parameters, like the magnitude and direction of the ferroelectric polarization and the thickness of the building blocks.  Also, the behavior of both electronic and structural relaxations at the polar/nonpolar interface needs to be better understood. Before addressing these aspects, we first focus on the polar/nonpolar interface only. Recently, promising 2DEG formation at the interface of the polar perovskite LaInO$_{3}$ and nonpolar perovskite BaSnO$_{3}$, has been reported by some of us~\cite{agg+21npj}. This heterostructure is motivated by the extraordinary room-temperature mobilities achieved in cubic BaSnO$_{3}$, reaching 320 cm$^{2}$(Vs)$^{-1}$~\cite{Hkim+12ape,paudel+17prb}. Thus, it can overcome the limitations of SrTiO$_{3}$ that exhibits about two orders of magnitude smaller mobilities~\cite{verma+2014PRL}. This material combination is also motivated by an experimental realization of a coherent interface with the lattice-matched perovskite LaInO$_{3}$ along the [001] direction~\cite{Martina+20prm}. The (LaO)$^+$ / (SnO$_2$)$^0$ interface was found to be most stable, suggesting the formation of a 2DEG~\cite{agg+21npj}. The latter work points to an important role of structural distortions in terms of octahedral tilts within LaInO$_{3}$ that affect the 2DEG charge density. As such, the LaInO$_{3}$/BaSnO$_{3}$ heterostructure (termed LIO/BSO from now on) is an ideal platform to investigate how to tune the characteristics of the interfacial 2DEG using ferroelectric layers. 

As the ferroelectric component we consider the perovskite BaTiO$_3$. At room temperature, it has a tetragonal structure with the ferroelectric polarization oriented along the [001] direction. Our choice is motivated by this characteristic which makes BaTiO$_3$ ideal to be combined with the LIO/BSO [001] heterostructure~\cite{agg+21npj}, offering the possibility of adjusting the polarization either toward or outward the interface. Therefore, the BaTiO$_3$/LaInO$_{3}$/BaSnO$_{3}$ combination (termed BTO/LIO/BSO from now on) exhibits all the key features to explore how to tune 2DEG properties using a ferroelectric material. 
\begin{figure*}[ht]
 \begin{center}
\includegraphics[width=.9\textwidth]{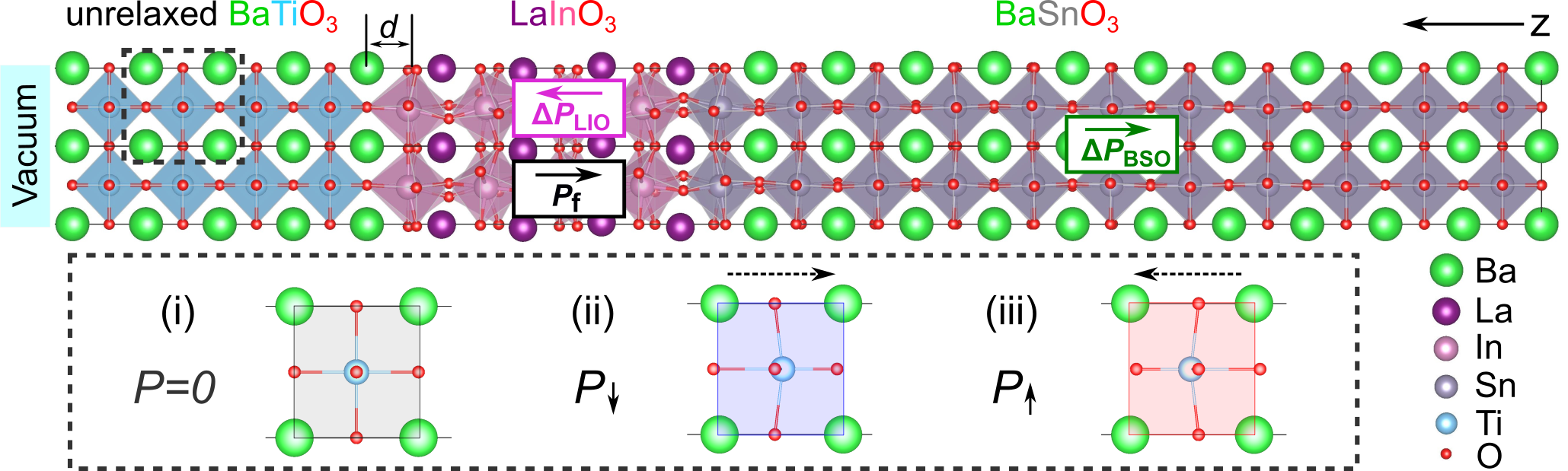}%
\caption{Sketch of the non-periodic BTO/LIO/BSO heterostructure (top) along the out-of-plane [001] direction (\textit{z}). $P_{\mathrm{f}}$ is the formal polarization within the LaInO$_{3}$ block oriented from the (InO$_{2}$)$^{-1}$ to the (LaO)$^{+1}$ plane. $\Delta P_{\mathrm{LIO}}$ (magenta) and $\Delta P_{\mathrm{BSO}}$ (green) indicates the polar distortions within the LaInO$_{3}$ and BaSnO$_{3}$ blocks. The distance $d$ separates LaInO$_{3}$ from BaTiO$_3$ that is fixed in all three cases, (i),(ii),(iii). The paraelectric case corresponds to zero polarization, \ie the centrosymmetric BaTiO$_3$ structure. Polarization toward ($P_{\downarrow}$) or outward ($P_{\uparrow}$) the LIO/BSO interface is induced by distorting atomic positions. }
\label{Fig-geometry}
 \end{center}
\end{figure*}

In this work, we comprehend how both density and distribution of the 2DEG formed at the LIO/BSO interface can be tuned by BaTiO$_3$, employing density-functional-theory (DFT). First, we present the characteristics of the pure LIO/BSO heterostructure. Then, we include the BaTiO$_3$ building block and consider three different scenarios which are (i) the paraelectric case ($P$=0), (ii) polarization toward of the LIO/BSO interface (termed $P_{\downarrow}$), and (iii) polarization outward the interface (termed $P_{\uparrow}$). We discuss the impact of orientation and magnitude of the polarization on the density and distribution of the 2DEG formed at the BaSnO$_{3}$ side. Finally, we focus on effects of structural relaxation in LIO/BSO induced by the ferroelectric BaTiO$_3$ layer and conclude how both electronic and structural relaxations can be tuned by controlling the polarization within the BaTiO$_3$ block. Overall, our results demonstrate the potential of this material combination for achieving high 2DEGs at polar/nonpolar interfaces and manipulating them. The most striking features of our work have been published recently~\cite{Fang2022}. Here, we provide an in-depth analysis behind our findings.

\section{Computational method}
\label{sec:method}
Ground-state properties are calculated using density-functional theory (DFT)~\cite{hohe-kohn64pr,kohn-sham65pr}, within the generalized gradient approximation (GGA) in the PBEsol parameterization~\cite{PBEsol+08prl} for exchange-correlation effects. All calculations are performed using FHI-aims~\cite{FHI-aims}, an all-electron full-potential package, employing numerical atom-centered orbitals. For all atomic species we use a \textit{tight} setting with a \textit{tier}-2 basis for oxygen (O), \textit{tier1+fg} for barium (Ba), \textit{tier1+gpfd} for tin (Sn), \textit{tier 1+hfdg} for lanthanum (La), \textit{tier 1+gpfhf} for indium (In) and \textit{tier} 1 for titanium (Ti). The convergence criteria are 10$^{-6}$ electrons for the density, 10$^{-6}$ eV for the total energy, 10$^{-4}$ eV/\AA~for the forces, and 10$^{-4}$ eV for the eigenvalues. For bulk BaTiO$_3$, lattice constants and internal coordinates are relaxed until the residual force on each atom is less than 0.001 eV/\AA. The sampling of the Brillouin zone is performed with an 8 $\times$ 8 $\times$ 8 $\textbf{k}$-grid. These parameters ensure converged total energies and lattice constants of 8 meV/atom and 0.001\AA,~respectively. The optimized bulk structures of BaSnO$_3$ and LaInO$_3$ are taken from Ref.~\cite{agg+21npj}.

For the heterostructures, the in-plane lattice parameters are fixed to $\sqrt{2}a_{\mathrm{BSO}}$ where $a_{\mathrm{BSO}}$ is the BaSnO$_3$ bulk lattice spacing. All heterostructures considered here are non periodic where a vacuum layer of about 140\,\AA\ is included and a dipole correction is applied in the [001] direction in order to prevent unphysical interactions between neighboring replica. We fix the first two BaSnO$_3$ unit cells (away from the interface) to the bulk geometry to simulate the bulk-like interior of the substrate. For the selected LIO/BSO interface, we first optimize the internal coordinates until the residual force on each atom is less than 0.001 eV/\AA. The respective geometry is then used to build the BTO/LIO/BSO heterostructure. The internal coordinates within the BaTiO$_3$ block are fixed in all considered cases. We change the atomic positions such to induce the desired polarization in terms of magnitude and direction (paraelectric, $P_{\uparrow}$, or $P_{\downarrow}$ case). The latter characteristics are controlled by tuning the relative cation-anion displacements. To analyze the role of structural relaxation inside the LIO/BSO building block for a given ferroelectric polarization, we perform a second optimization of its internal coordinates in the presence of the fixed BaTiO$_3$ component. For the selfconsistent calculations, a 6 $\times$ 6 $\times$ 1 $\textbf{k}$-grid is adopted for all systems, for the electronic properties, a 20 $\times$ 20 $\times$ 1 $\textbf{k}$-grid. This parameter ensures converged and electron/hole charge densities within 0.01 $e/a^{2}$. 
\begin{figure*}[htp]
 \begin{center}
\includegraphics[width=.9\textwidth]{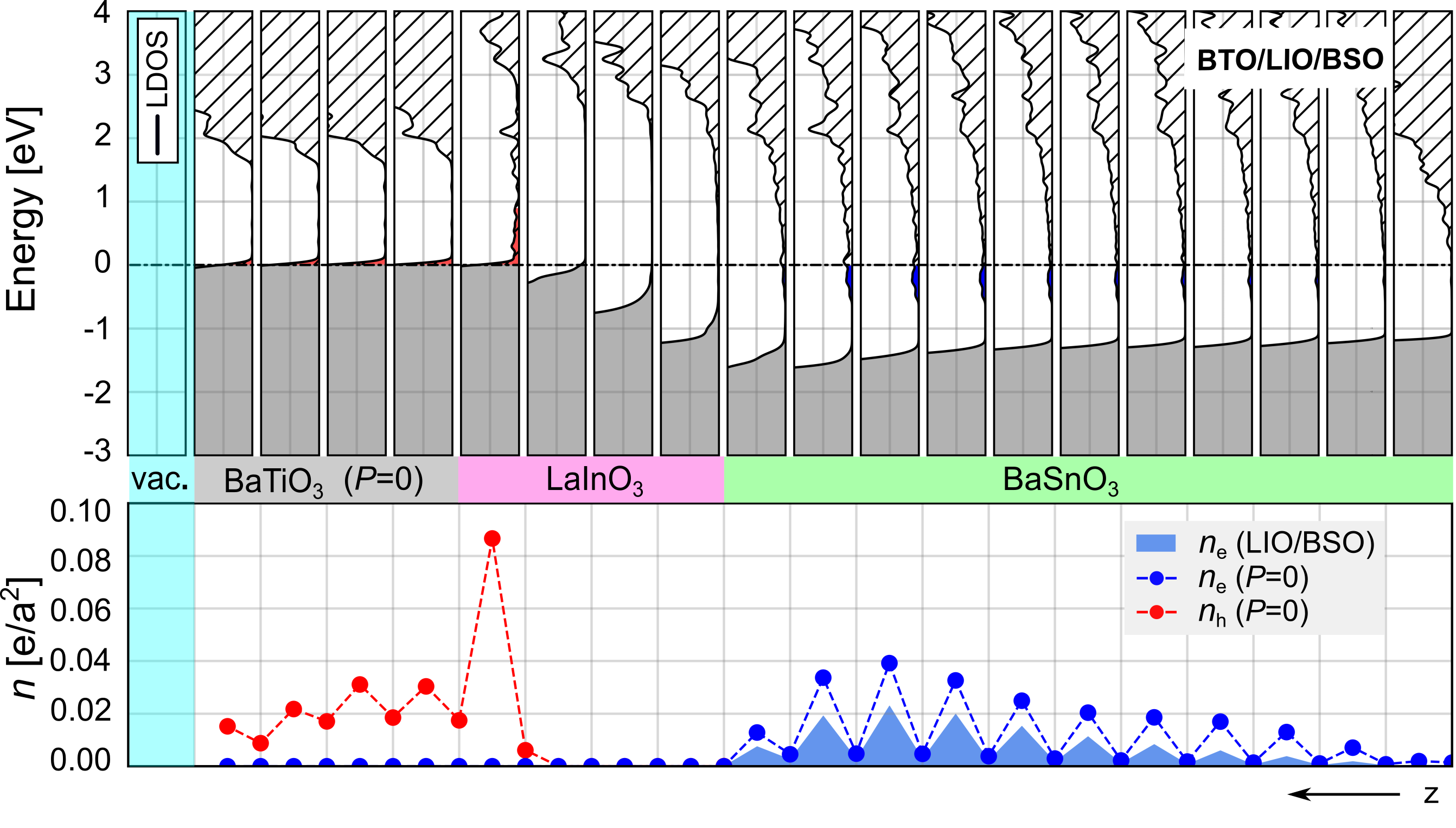}
\caption{Electronic properties of the BTO/LIO/BSO heterostructure in the paraelectric case. In the LDOS (top), the gray area indicates the occupied valence states (Fermi level at zero). The depleted valence-band region (holes) and the occupied conduction-band region (electrons) resulting from electronic reconstruction are marked by red and blue color, respectively. The panel below depicts the distribution of the electron (hole) charge density.   } 
 \label{Fig-para}
 \end{center}
\end{figure*}

The Born effective charges of the pristine materials, Z$^{*}$, are computed within the Berry-phase approach~\cite{berryPhase+93prbr} using \texttt{exciting}~\cite{gula+14jpcm}, an all-electron full-potential code, implementing the family of (L)APW+LO (linearized augmented planewave plus local orbital) methods. These calculations are performed for the bulk structure of BaTiO$_{3}$ as obtained by the FHI-aims code, using the same $\textbf{k}$-grid. For the atomic species Ba, Ti, and O, muffin-tin radii (R$_{MT}$) of 2.0, 1.8 and 1.6 bohr are used, respectively. A basis-set cutoff R$_{MT}$G$_{max}$=7 is adopted, where R$_{MT}$ here refers to the radius of the smallest sphere (1.6 bohr), {\it i.e.}, G$_{max}$= 4.375 bohr$^{-1}$. Atomic structures are visualized using the software package VESTA~\cite{momm-izum11jacr}.

\section{Pristine materials}
\label{pristine}
Before proceeding with the heterostructures, we summarize the structural and electronic properties of the pristine perovskites BaSnO$_3$ and LaInO$_3$ that are detailed in Refs.~\cite{agg+21npj,Aggoune+LIO,Aggoune+BSO}. BaSnO$_3$ exhibits a cubic (Pm$\overline 3$m) perovskite structure with a calculated lattice parameter of 4.119~\AA. LaInO$_3$ has an orthorhombic (Pbnm) structure containing four formula units per unit cell. The optimized structural parameters \textit{a}= 5.70~\AA, \textit{b}= 5.94~\AA, and \textit{c}= 8.21~\AA\ are in good agreement with experimental values~\cite{zbigniew+20pssb}. The lattice parameter of the corresponding pseudocubic unit cell of LaInO$_3$ is 4.116~\AA~\cite{agg+21npj}, implying a very small lattice mismatch with BaSnO$_3$ of about 0.07$\%$.
\begin{figure*}[htp]
 \begin{center}
\includegraphics[width=.9\textwidth]{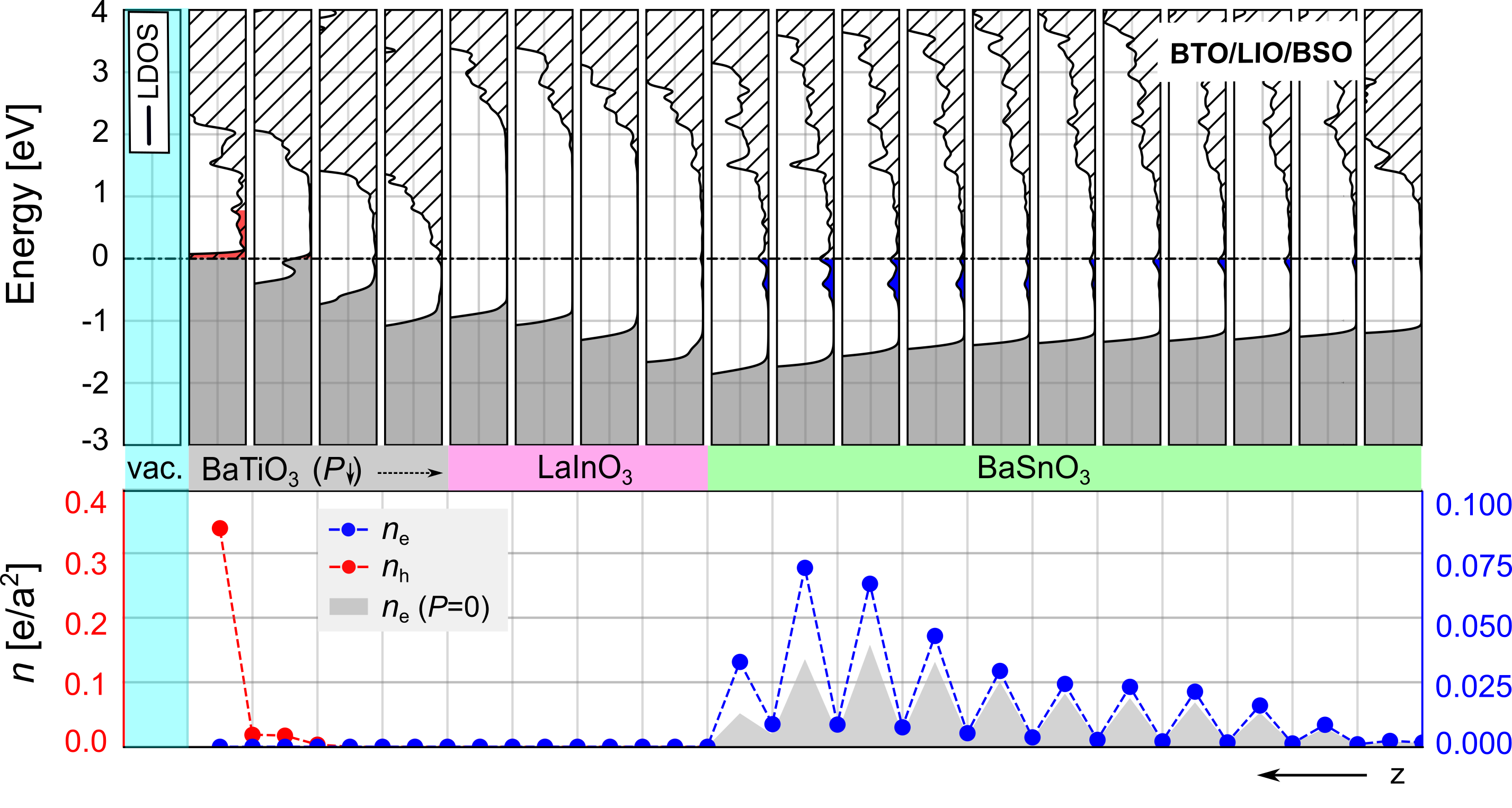}%
\caption{Electronic properties of the BTO/LIO/BSO heterostructure with a ferroelectric polarization of $P_{\downarrow}$=--0.12 C/m$^2$, \ie oriented towards the interface. The top panel shows the LDOS, where the gray area marks the occupied valence states (Fermi level set to zero). The depleted valence-band region (holes) and the occupied conduction-band region (electrons) resulting from electronic reconstruction are indicated by red and blue color, respectively. The bottom panel depicts the distribution of the electron (hole) charge density.} 
\label{Fig-pdown}
 \end{center}
\end{figure*}

The ferroelectric material BaTiO$_3$ has a tetragonal phase (P4mm) at room temperature, with an experimental lattice constant of $a_{\textrm{BTO}}$=3.990~\AA~and a $c/a$ ratio of 1.011 \cite{li+1998Asc}. The measured band gap ranges between 3.2 and 3.5 eV \cite{wemple+1970PRB,zen+2013APL}, the ferroelectric polarization was determined to be 0.27 C/m$^2$ and oriented along the [001] direction~\cite{zhong+1994PRL}. Previous calculations report a lattice constant $a$=3.977\,\AA, a $c/a$ ratio of 1.022, and a ferroelectric [001] polarization of 0.31 C/m$^2$~\cite{wu+2016PRB}. The Kohn-Sham band gap, obtained by applying a Hubbard correction \cite{pet+2003PRB,lie+1995PRB,per+1996PRL} to the Ti-\textit{d} states computed with the PBE functional, is about 2 eV\,\cite{gao+2016PE}, underestimating the experimental counterpart by $\sim$40$\%$~\cite{wemple+1970PRB,zen+2013APL}. In our work, the optimized lattice parameters ($a$=3.954\,\AA, $c/a$=1.023) obtained with PBEsol is slightly underestimated. The Kohn-Sham band gap is calculated to be 1.82 eV, close to a PBE value of 1.73\,eV \cite{zhang+2017PRB}. The calculated Born effective charges, Z$^{*}_{z,i}$, along the \textit{z} direction (c axis) are 2.86 for Ba, 5.71 for Ti, -4.64 for the O atom within the BaO plane and -1.96 for that within the TiO$_2$ plane. These values are also in good agreement with a previous theoretical report \cite{hermet+2009JPCM}. The polarization is given as
\begin{align}
P_z= \frac{1}{\Omega} \sum_{i} Z^{*}_{z,i}\Delta_{z,i}
\end{align}
where $\Omega$ is the unit-cell volume and $\Delta_{z,i}$ is the relative out-of-plane displacement of the  $i$-th cation Ba (Ti) with respect to the anion O within the BaO (TiO$_2$) plane. The estimated \textit{z}-axis polarization of BaTiO$_3$ is about 0.29 C/m$^2$ which matches very well with the experimental counterpart~\cite{zhong+1994PRL}.

As mentioned in Section \ref{sec:method}, the in-plane lattice parameters of the BTO/LIO/BSO heterostructures are fixed to that of bulk BaSnO$_{3}$ which is considered as the substrate. As such, the BaTiO$_{3}$ building block is strained. Therefore, before building the heterostructures, we present the properties of the constrained BaTiO$_{3}$. For the in-plane lattice parameter of 4.119~\AA, we find a smaller tetragonal distortion, \ie $c/a$=0.940 and a decrease of the Kohn-Sham band gap to 1.59\,eV. The ferroelectric [001] polarization decreases to about 0.10\,C/m$^2$.

\section{2DEG at the LIO/BSO interface}
We start our analysis with a non-periodic LIO/BSO heterostructure, consisting of a thin LaInO$_{3}$ layer on top of a [001] BaSnO$_{3}$ substrate. We consider a stoichiometric system and focus on the (LaO)$^{+1}$/(SnO$_2$)$^0$ interface that is supposed to be most favorable~\cite{agg+21npj}. Here, BaSnO$_{3}$ has a thickness of 11 unit cells, which can capture the extension of the structural deformations and the 2DEG distribution away from the interface~\cite{agg+21npj}. The LaInO$_{3}$ block spans four pseudocubic unit cells and is terminated with an (InO$_{2})^{-1}$ plane at the surface. The formal polarization, $P_{\mathrm{f}}$, that points from the surface toward the interface, amounts to 0.47 C/m$^2$ and leads to a polar discontinuity at the interface with the nonpolar BaSnO$_{3}$. The structural model of this system is depicted at the bottom of Fig.~\ref{Fig-1}, where we show a compilation of its electronic properties. 

Before discussing the motivation behind the choice of the LaInO$_{3}$ thickness, we note that structural and electronic properties of LIO/BSO interfaces have been discussed in detail by some of us in Ref.~\onlinecite{agg+21npj}. In this work, an electronic reconstruction to compensate the interfacial polar discontinuity was reported, giving rise to the formation of a 2DEG at the interface. It exhibits Sn-\textit{s} character and is confined within the BaSnO$_{3}$ side, while the corresponding two-dimensional hole gas (2DHG) at the surface is hosted by O-\textit{p} states within the (InO$_{2})^{-1}$ plane. The 2DEG density was found to gradually increase with LaInO$_{3}$ thickness. This behavior is governed by structural distortions within LaInO$_{3}$ which induce a depolarization field to compensate the polar discontinuity at the LIO/BSO interface (termed polar distortion~\cite{agg+21npj}). The spatial distribution of the 2DEG was shown to be tunable by changing the dimensionality of the system (\textit{i.e.} considering either periodic or non-periodic interface)~\cite{agg+21npj}. 

As we target to control, \ie also enhance the 2DEG density and distribution, it is appropriate to consider an interface that has low 2DEG density. For this reason, the LaInO$_{3}$ block, considered here, has a thickness of four pseudocubic unit cells, which generates a 2DEG with a density of about 0.13 $e/a^2$~\cite{agg+21npj}. As evident from Fig.~\ref{Fig-1}(a), the electronic reconstruction is caused by the dipole induced within the polar material LaInO$_{3}$ which leads to an upward shift of the valence-band edge. At the surface, the valence-band edge crosses the Fermi level, leading to a charge transfer to the interface. Consequently, the bottom of the conduction band becomes partially occupied within the BaSnO$_{3}$ side, giving rise to a 2DEG that is confined within five unit cells ($\sim$20\AA). Consequently at the surface, a 2DHG is formed that is confined within one LaInO$_{3}$ pseudocubic unit cell. To partially compensate the polar discontinuity, polar distortions (termed $\Delta P$) appear at the interface, analogous to what was described above ~\cite{agg+21npj}. It is oriented opposite (parallel) to the formal polarization $P_{\mathrm{f}}$ within the LaInO$_{3}$ (BaSnO$_{3}$) block. As such, it weakens the electronic reconstruction. We note that despite the underestimation of the band gaps of the pristine materials by PBEsol, this functional is good enough to capture the band alignment across the LIO/BSO interface as well as both density and spacial distribution of the 2DEG~\cite{agg+21npj}. Overall, the characteristics of the selected interface are ideal to investigate the impact of a ferroelectric BaTiO$_{3}$ layer. \\

\section{Impact of the ferroelectric material}
\label{sec:unrelaxed}

We proceed our analysis by investigating the electronic properties of the BTO/LIO/BSO [001] heterostructures. We consider three scenarios: (i) the paraelectric case ($P$=0), (ii) polarization toward the LIO/BSO interface ($P_{\downarrow}$), and (iii) polarization outward of the interface ($P_{\uparrow}$). These scenarios are realized by atomic displacements within the BaTiO$_3$ block (see Section~\ref{sec:method} and Fig.~\ref{Fig-geometry}) which allows also to control the polarization magnitude. This approach mimics the experimental situation where the ferroelectric polarization (orientation and magnitude) is controlled by applying an electric voltage~\cite{kim+2013AM}. In a first step, we combine the BaTiO$_3$ layers with the LIO/BSO interface without further relaxation. In this way, we focus on how both orientation and magnitude of the ferroelectric polarization impact the electronic reconstruction only. Nevertheless, we determine the distance between BaTiO$_3$ and LaInO$_{3}$, \textit{i.e.} between the adjacent (InO$_2$)$^{-1}$ and (BaO)$^0$ layers to be $d$=2.2~\AA. 

Regarding the thickness of the BaTiO$_3$ block, we note that even if epitaxial layers used experimentally are typically tens of nanometers thick or even more~\cite{kim+2013AM,wang+2018ACS,gao+2018AMI}, our choice of four unit cells ($\sim$16~\AA, Fig.~\ref{Fig-geometry}) is justified as it allows us to tune the magnitude of the ferroelectric polarization gradually and capture its effect on the LIO/BSO interface. It also allows to investigate two regimes, with and without a band gap in the BaTiO$_3$ building block. With these considerations, we now analyze the electronic properties of the BTO/LIO/BSO heterostructures. 
\begin{figure}[h]
\begin{center}
\includegraphics[width=.45\textwidth]{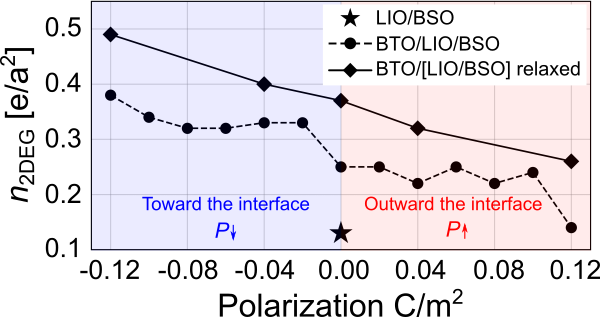}%
\caption{2DEG densities within the BaSnO$_3$ block of the BTO/LIO/BSO heterostructures as a function of ferroelectric polarization. $P$=0 refers to the paraelectric case, negative (positive) values to $P_{\downarrow}$ ($P_{\uparrow}$). Circles (diamonds) are results for heterostructures without (with) further relaxation of the LIO/BSO interface with the BaTiO$_3$ block kept fixed. The value of the pristine LIO/BSO interface is indicated by the star.}
\label{Fig-2DEG-vs-P}
\end{center}
\end{figure}

\subsection{Paraelectric BaTiO$_{3}$}

The description of a paraelectric BaTiO$_3$ building block --no ferroelectric polarization-- is achieved by considering its centrosymmetric structure. This way, we analyze how a bare BaTiO$_3$ layer impacts the characteristics of the LIO/BSO interface. As depicted in Fig.~\ref{Fig-para}, the LDOS is very similar to that of the pristine LIO/BSO interface shown in Fig.~\ref{Fig-1}. Integrating it over the partially filled states within the BaSnO$_3$ block, we find a similar 2DEG distribution but with higher density, compared to the single LIO/BSO interface.

The enhanced electronic charge is attributed to the fact that the valence-band edge at the BaTiO$_{3}$ side of the BTO/LIO interface crosses the Fermi level (as it does at the LaInO$_{3}$ side) leading to partially unoccupied states. This, in turn, enhances the electronic charge transfer to the BaSnO$_{3}$ side, increasing the 2DEG density. The confined 2DHG within the LaInO$_{3}$ side is consequently extended into the BaTiO$_{3}$ layers. To summarize, the electronic charge density within BaSnO$_{3}$ (and likewise the 2DHG density) is nearly doubled, \ie increases to about 0.25 $e/a^2$ compared to 0.13 $e/a^2$ in the pristine LIO/BSO interface. In the following, we consider this configuration as a reference to investigate the effect of ferroelectric polarization within Section~\ref{sec:unrelaxed}.
\begin{figure*}[hbt]
 \begin{center}
\includegraphics[width=.9\textwidth]{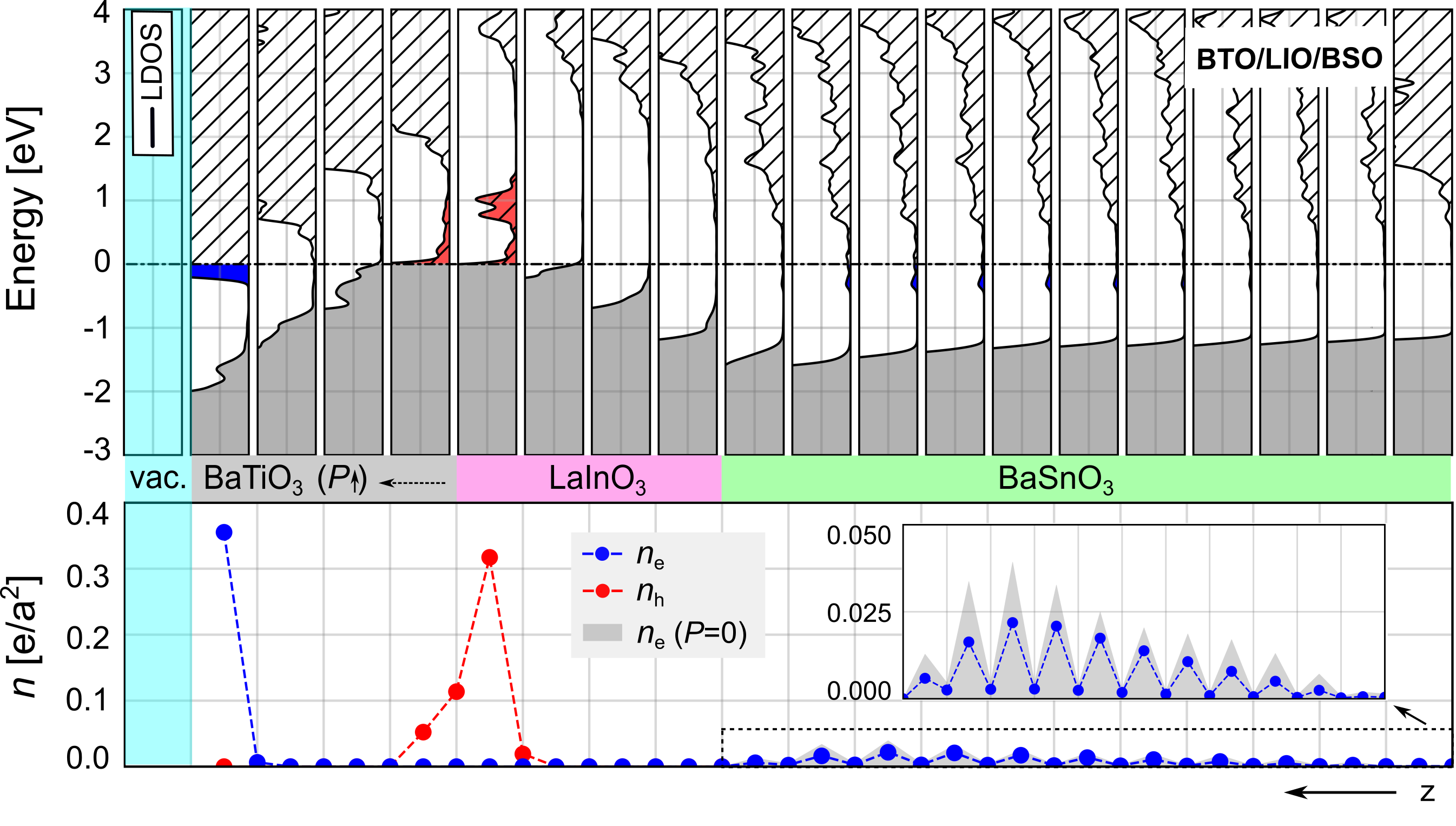}%
\caption{Same as Fig. \ref {Fig-pdown} but with a ferroelectric polarization of +0.12 C/m$^2$ ($P_{\uparrow}$). For better visibility, the inset zooms into the 2DEG density within the BaSnO$_3$ side.}
\label{Fig-up}
 \end{center}
\end{figure*}

\subsection{Polarization toward the interface}
To start with, we choose a ferroelectric BaTiO$_3$ polarization of $P_{\downarrow}$=--0.12 C/m$^2$, where the negative sign indicates that it points opposite to the \textit{z} direction. As we see in Fig.~\ref{Fig-pdown}, the dipole induced within the BaTiO$_3$ block causes an upward shift of the valence-band edge, starting from the LaInO$_3$ side and being most pronounced at the BaTiO$_3$ surface. This effect is also visible in the electrostatic potential, depicted in Fig.~S1 of the Supporting Information (SI). Consequently, the hole charge increases and localizes at the BaTiO$_3$ surface rather than at the BTO/LIO interface. This, in turn, amplifies the electronic charge density within the BaSnO$_3$ block that reaches a value of about 0.38 $e/a^2$. Overall, this polarization direction, being parallel to $P_{\mathrm{f}}$, enhances the polar discontinuity which, in turn, enhances the 2DEG density. From an electrostatic point of view, the electronic charge needs to localize as close as possible to the interface for an effective compensation of the polar discontinuity~\cite{Bristowe+09prb,Stengel+11prl}. For this reason, enforcing the polar discontinuity by BaTiO$_3$, leads to an accumulation of the 2DEG near the LIO/BSO interface (Fig.~\ref{Fig-pdown}). This explains the possibility of enhancing and confining the 2DEG by a ferroelectric layer.

To understand these characteristics, we vary the ferroelectric polarization between 0 and --0.12 C/m$^2$. Figure~\ref{Fig-2DEG-vs-P} shows that the charge density increases by increasing magnitude of polarization. This trend indicates that one can reach higher 2DEG densities by increasing the polarization magnitude. 

\subsection{Polarization against the interface}
To complete the picture, we switch the polarization direction, now pointing outward the LIO/BSO interface (termed $P_{\uparrow}$). The results for a magnitude of +0.12 C/m$^{2}$ are depicted in Fig.~\ref{Fig-up}. In contrast to the previous case, \ie $P_{\downarrow}$, the dipole induced within the BaTiO$_3$ block causes an upward shift of the valence-band edge, starting at the BaTiO$_3$ surface and being most pronounced at the BTO/LIO interface. (For the effect on the electrostatic potential, see Fig.~S1 in the SI.) Consequently, the hole charge increases and localizes at the BTO/LIO interface which, in turn, enhances the electronic charge density. However, we find that a part of the charge is transferred to the BaTiO$_3$ surface to compensate its ferroelectric dipole. This process drastically reduces the 2DEG density within the BaSnO$_3$ block to about 0.14 $e/a^2$ compared to the paraelectric case (0.25 $e/a^2$) (see Figs.~\ref{Fig-2DEG-vs-P} and ~\ref{Fig-up}). For completing the study, we vary the ferroelectric from 0 to +0.12 C/m$^2$. As we can see in Fig.~\ref{Fig-2DEG-vs-P}, the charge density within the BaSnO$_3$ side decreases by increasing polarization magnitude. Overall, this scenario offers the possibility of depleting the 2DEG charge.
\begin{figure*}
 \begin{center}
\includegraphics[width=.8\textwidth]{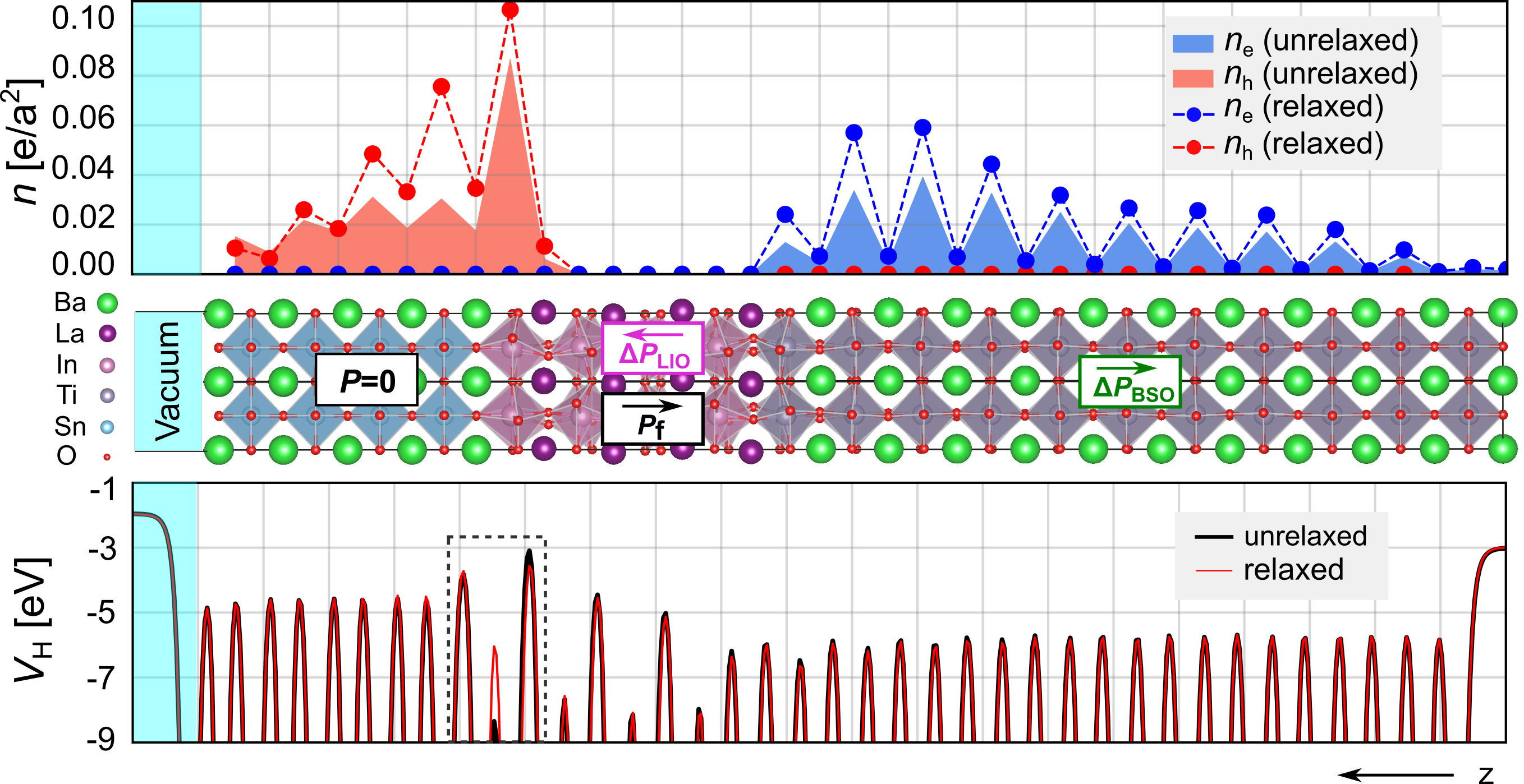}%
\caption{Sketch of the BTO/LIO/BSO heterostructure where the LIO/BSO component is relaxed at fixed BaTiO$_3$ geometry. $P_{\mathrm{f}}$ is the formal polarization oriented from the (InO$_{2}$)$^{-1}$ plane towards the (LaO)$^{+1}$ plane at the interface. $\Delta P_{\mathrm{LIO}}$ ($\Delta P_{\mathrm{BSO}}$) is the polarization due to structural distortions within the LaInO$_{3}$ (BaSnO$_{3}$) side. The top panel shows the distribution of electron and hole charge densities compared to the unrelaxed situation in the paraelectic case. In the bottom panel, the in-plane averaged electrostatic potential is depicted. The dashed rectangle highlights the region with pronounced difference.}
\label{Fig-unrelaxed-vs-relaxed}
 \end{center}
\end{figure*}

\section{Structural relaxation effects}
To validate our findings and analyze how possible structural distortions impact the resulting 2DEG density, we optimize the BaSnO$_3$ and LaInO$_3$ blocks of the BTO/LIO/BSO heterostructures, keeping the BaTiO$_3$ component fixed. We first consider the paraelectric case. As depicted in Fig.~\ref{Fig-unrelaxed-vs-relaxed}, the 2DEG (and likewise the 2DHG) density is enhanced, reaching a value of 0.37 $e/a^2$ that is 0.12 $e/a^2$ higher compared to the unrelaxed case (see also \textit{P}=0 in Fig.~\ref{Fig-2DEG-vs-P}). To understand this behavior, we look at the in-plane averaged electrostatic potential that is depicted in the bottom of the figure. We observe an upward shift right at the BTO/LIO interface, particularly in the closest InO$_{2}$ layer. This can be understood as this layer is no longer at the surface but now forms the interface to the ferroelectric material. Upon relaxation, its presence reduces the depolarization field ($\Delta P_{\mathrm{LIO}}$) induced by the structural distortions within the polar LaInO$_3$ block (Figs.~\ref{Fig-1} and \ref{Fig-unrelaxed-vs-relaxed}). Thus, the dipole induced within LaInO$_3$ (mainly due to the formal polarization $P_{\mathrm{f}}$), is enhanced, making the polar discontinuity at the LIO/BSO interface more pronounced. As such, the 2DEG (and likewise the 2DHG) density increases in order compensate it. Overall, this analysis indicates that structural relaxations at the LIO/BSO interface in the presence of the BaTiO$_3$ block enhance the 2DEG density even without ferroelectric polarization ($P$=0). In the following, the charge density and distribution of the relaxed paraelectric case will be considered as a reference.

\begin{figure*}
 \begin{center}
\includegraphics[width=.8\textwidth]{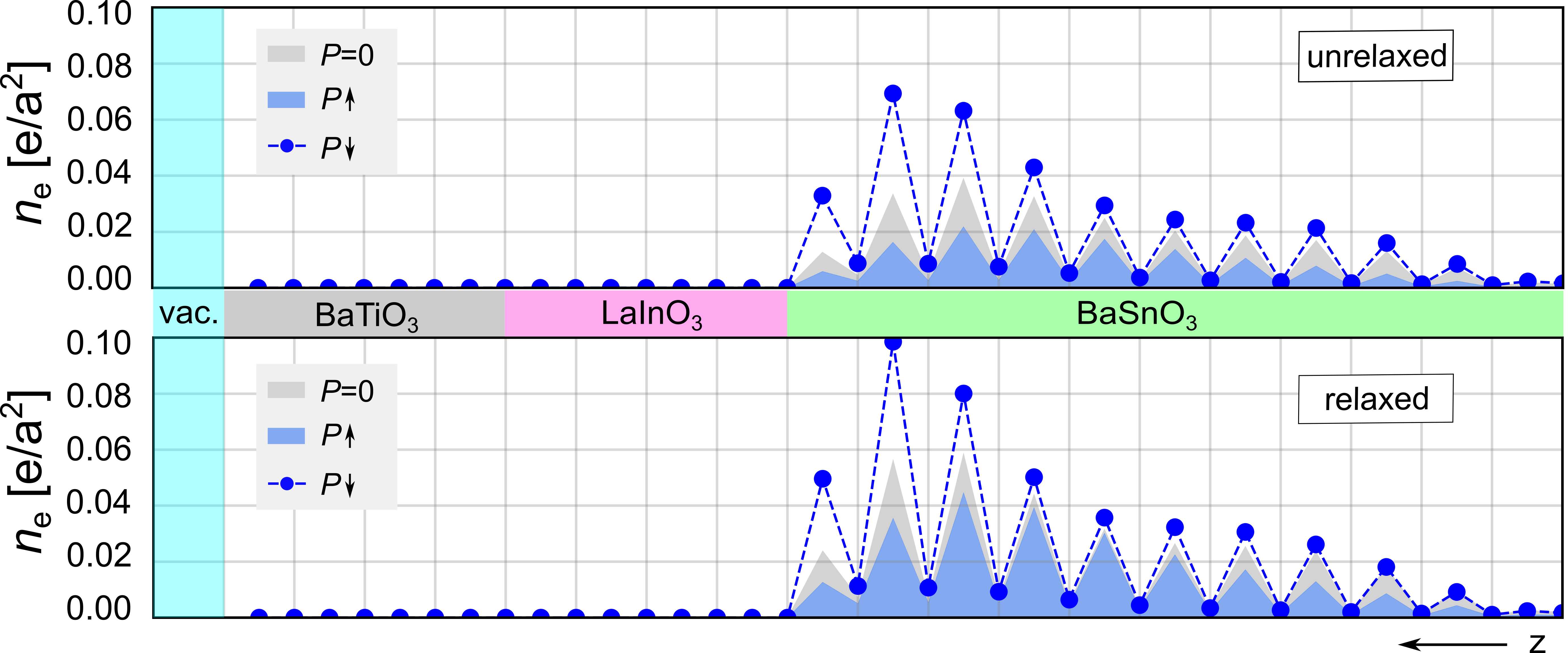}%
\caption{Charge-density distribution within the BaSnO$_3$ component of the BTO/LIO/BSO heterostructure for the unrelaxed (top) and relaxed (bottom) LIO/BSO interface.}
\label{Fig-relaxed-up-down}
 \end{center}
\end{figure*}

Now we relax the LIO/BSO block at a polarization of $\pm$0.12 C/m$^2$. The results are shown in Fig.~\ref{Fig-relaxed-up-down}. We see that the spatial extent of the charge distribution within the BaSnO$_3$ block hardly changes, but its magnitude decreases. For details on the LDOS and the hole distribution, we refer to Figs.~S3 and S4 in the SI. Compared to the paraelectric case, we clearly see that, independent of the relaxation, $P_{\downarrow}$ causes an accumulation of electronic charge across the LIO/BSO interface while, $P_{\uparrow}$ results in charge depletion. As such, this finding validates the behavior found with the unrelaxed structures. Upon relaxation, the charge density reaches a value of 0.49 $e/a^2$ (0.26 $e/a^2$) for the $P_{\downarrow}$ ($P_{\uparrow}$) case. Finally, we consider  ferroelectric polarizations of $\pm$0.04 C/m$^2$ (Fig.~\ref{Fig-2DEG-vs-P}) which confirm the linear trend of the charge density as a function of the polarization magnitude. 

Before concluding, we like to discuss how our predictions based on ideal structures can be useful for practical applications. As epitaxial ferroelectric layers used experimentally are usually much thicker (tens of nanometers or more)~\cite{kim+2013AM,wang+2018ACS,gao+2018AMI}, we perform additional calculations with varying thickness of the BaTiO$_{3}$ block. Decreasing it to three unit cells and setting the ferroelectric polarization to $P_{\downarrow}$=--0.05 C/m$^2$, corresponds to  four unit cells and $P_{\downarrow}$=--0.04 C/m$^{2}$. In this way, we fix the induced electric field and  can analyze the role of the BaTiO$_{3}$ thickness independently (see SI). We find that the charge density increases with the thickness (Fig. S2 in the SI) and conclude that we can expect a further enhancement in real samples. Since it shows a linear trend as a function of the ferroelectric polarization, we expect the 2DEG density to be effectively tunable in real samples where stronger polarization can be induced. 

\section{Conclusions}

In summary, we have presented a detailed first-principles study on how to control the characteristics of a 2DEG at a polar/nonpolar interface by adding a ferroelectric functional layer. To this extent, we have considered LaInO$_3$/BaSnO$_3$ as a prototypical interface. In this system, a 2DEG (2DHG) with a density of about 0.13 $e/a^2$ is formed across the interface (at the surface). Building a BaTiO$_3$/LaInO$_3$/BaSnO$_3$ heterostructure, the ferroelectric component BaTiO$_3$ induces fascinating effects on the 2DEG and 2DHG charges. Interestingly, their densities increase to  0.25 $e/a^2$ already by adding a paraelectric BaTiO$_3$ block ($P$=0). A ferroelectric polarization of +0.12 C/m$^{2}$ toward the LaInO$_3$/BaSnO$_3$ interface, accumulates charge within BaSnO$_3$, enhancing the 2DEG density to about 0.38 $e/a^2$. Thereby, the 2DHG distributes over the BaTiO$_3$ surface. Switching the polarization direction to pointing outward the LaInO$_3$/BaSnO$_3$ interface, depletes the 2DEG charge to 0.14 $e/a^2$. This charge transfer into the ferroelectric block appears to compensate its local dipole, forming another 2DEG but at the BaTiO$_3$ surface. Our results indicate the possibility of tuning the charge density within the BaSnO$_3$ substrate by switching the ferroelectric polarization. Optimizing the LaInO$_3$/BaSnO$_3$ interface at fixed BaTiO$_3$ geometry, leads to an enhancement of the 2DEG density. This characteristic is attributed to the fact that the presence of the BaTiO$_3$ layer reduces the structural distortions within the polar LaInO$_3$ block. Varying the magnitude of the ferroelectric polarization, changes the 2DEG density linearly. Our findings demonstrate and open up a perspective for controlling the properties of 2DEGs at emerging polar/nonpolar interfaces by ferroelectric polarization as an external stimulus.


\section*{Data availability}
Input and output files can be downloaded free of charge from the NOMAD Repository~\cite{drax-sche19jpm} at the following link: \url{https://dx.doi.org/10.17172/NOMAD/2022.01.11-1}. The data of the unrelaxed case can be found at this link:  \url{https://dx.doi.org/10.17172/NOMAD/2022.03.01-1}

\section*{Acknowledgment} 

L.F. and W.R. are grateful for support from the National Natural Science Foundation of China (51861145315, 11929401, 12074241), the Independent Research and Development Project of the State Key Laboratory of Advanced Special Steel, Shanghai Key Laboratory of Advanced Ferrometallurgy, Shanghai University (SKLASS 2020-Z07), the Science and Technology Commission of Shanghai Municipality (19DZ2270200, 19010500500, 20501130600), and the China Scholarship Council (CSC). This work was supported by the project BaStet (Leibniz Senatsausschuss Wettbewerb, No. K74/2017) and was performed in the framework of GraFOx, a Leibniz Science Campus, partially funded by the Leibniz Association. We acknowledge the North-German Supercomputing Alliance (HLRN) for providing HPC resources that have contributed to the research results reported in this paper (project bep00078 and bep00096). W.A. and L.F. thank Martin Albrecht, Martina Zupancic (Leibniz-Institut f\"{u}r Kristallz\"{u}chtung, Berlin), Dmitrii Nabok (Humboldt-Universit\"{a}t zu Berlin), Chen Chen (Shanghai University), and Kookrin Char (Seoul National University) for fruitful discussions. 

\section*{Additional information}
Supplementary information is available for this paper at (/https:/...).

\newpage

\renewcommand{\figurename}{}
\renewcommand{\tablename}{}
\setcounter{figure}{0}    
\renewcommand{\thepage}{S\arabic{page}} 
\renewcommand{\thetable}{Supplementary table \arabic{table}}  
\renewcommand{\thefigure}{Supplementary Figure \arabic{figure}} 

\newpage
\onecolumngrid

\newpage
{\centering
{\large 
\textbf{Supporting Information on}\\
\vskip 0.2cm
\textbf{"How ferroelectric BaTiO$_3$ can tune a two-dimensional electron gas at the interface of LaInO$_{3}$ and BaSnO$_{3}$: a first-principles study"}}\newline

Le Fang$^{1,2}$, Wahib Aggoune$^{2}$, Wei Ren$^{1,3}$, and Claudia Draxl$^{2,4}$\\
\textit{$^{1}$Materials Genome Institute, International Center for Quantum and Molecular Structures, Physics Department, Shanghai University, 200444 Shanghai, China}\\
\textit{$^{2}$Institut f\"{u}r Physik and IRIS Adlershof, Humboldt-Universit\"{a}t zu Berlin, 12489 Berlin, Germany}\\
\textit{$^{4}$Shanghai Key Laboratory of High Temperature Superconductors and State Key Laboratory of Advanced Special Steel, Shanghai University, Shanghai 200444, China}\\
\textit{$^{4}$European Theoretical Spectroscopy Facility (ETSF)}\\

\vskip 0.5cm
\date{today} 
}

 \twocolumngrid
   \normalsize 

\subsection{Heterostructure with unrelaxed $\mathrm{LaInO_{3}/BaSnO_{3}}$ interface}
\subsubsection{In-plane averaged electrostatic potential}
\begin{figure*}[hbt]
 \begin{center}
\includegraphics[width=.95\textwidth]{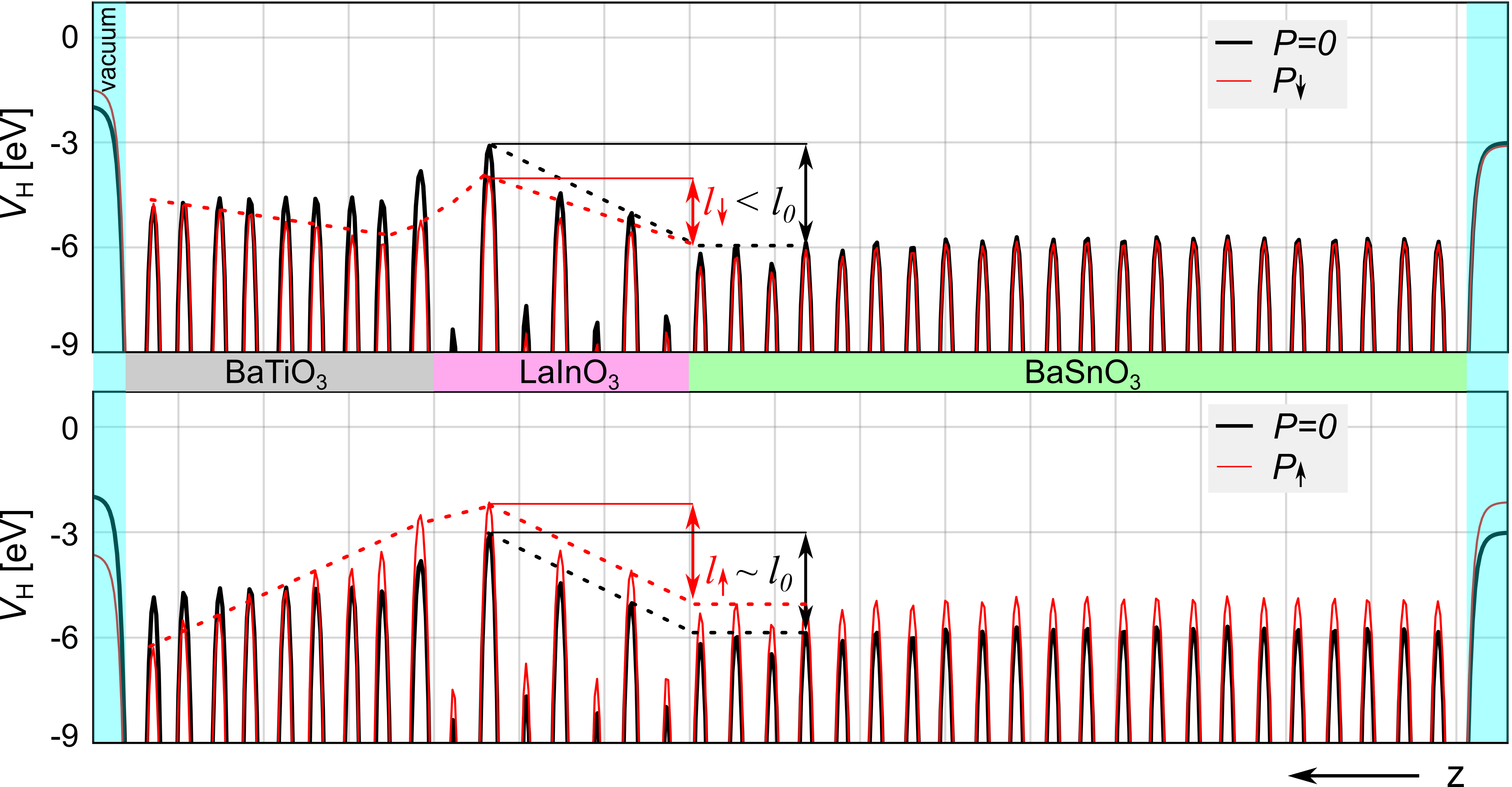}%
\renewcommand*{\thefigure}{S\arabic{figure}}
\caption{In-plane averaged electrostatic potential of the BTO/LIO/BSO heterostructure along the z direction for polarization toward the interface ($P_{\downarrow}$=--0.12 C/m$^2$, top panel) and in opposite direction ($P_{\uparrow}$=+0.12 C/m$^2$, bottom panel) compared to the paraelectric case (black line). The dashed lines are guides to the eye, highlighting the trend of the potential. The horizontal solid black line indicates the maximum of the potential within the LaInO$_{3}$ block for the praraelectric case, the respective red solid lines for $P_{\downarrow}$ and $P_{\uparrow}$, respectively. $l_0$, $l_{\downarrow}$, and $l_{\uparrow}$ indicate their differences to the respective values in the BaSnO$_{3}$ block. 
}
\label{Fig-potential}
 \end{center}
\end{figure*}
In the top panel of Fig.~\ref{Fig-potential}, we depict the in-plane averaged electrostatic potential of the BaTiO$_{3}$/LaInO$_{3}$/BaSnO$_{3}$ (BTO/LIO/BSO) heterostructure for the unrelaxed LIO/BSO interface, choosing a polarization of $P_{\downarrow}$=--0.12 C/m$^2$. The potential remains unchanged within the BaSnO$_{3}$ side compared to the paraelectric case, while it is lowered across the BTO/LIO interface, reflecting the induced dipole in the BaTiO$_3$ block. The increase of the potential (built-in potential) within LaInO$_3$ is a direct consequence of its polar character. Its magnitude (termed $l_0$, $l_{\downarrow}$, and $l_{\uparrow}$ for the paraelectric, $P_{\downarrow}$ and $P_{\uparrow}$ case, respectively) indicates the changes in polar discontinuity at the LIO/BSO interface due to the polarization of the BaTiO$_3$ block. There is a clear difference observed for the two polarization directions. For the $P_{\downarrow}$ case, we see that $l_{\downarrow}$~$\textless$~$l_0$ (top panel), \textit{i.e.}, the polar discontinuity at the LIO/BSO interface is reduced, which is consistent with the increased 2DEG density at the BaSnO$_{3}$ side. In this case, the hole gas localizes at the BaTiO$_3$ surface due to its electric dipole that leads to an upward shift of the valence-band maximum at the surface. For the opposite polarization ($P_{\uparrow}$=+0.12 C/m$^2$), the polar discontinuity at the LIO/BSO interface appears unaffected ($l_{\uparrow}$~$\sim$~$l_0$). However, as shown in Fig.~6 of the main text, the 2DEG density inside BaSnO$_3$ is depleted due to additional charge transfer from the BTO/LIO interface into the BaTiO$_{3}$ surface. This can be seen by the decrease of the potential (likewise the valence-band edge) within the BaTiO$_{3}$ block which is most pronounced at the surface, mainly due to its dipole. 

\subsubsection{Role of BaTiO$_3$ thickness}
\begin{figure*}[htbp]
 \begin{center}
\includegraphics[width=.95\textwidth]{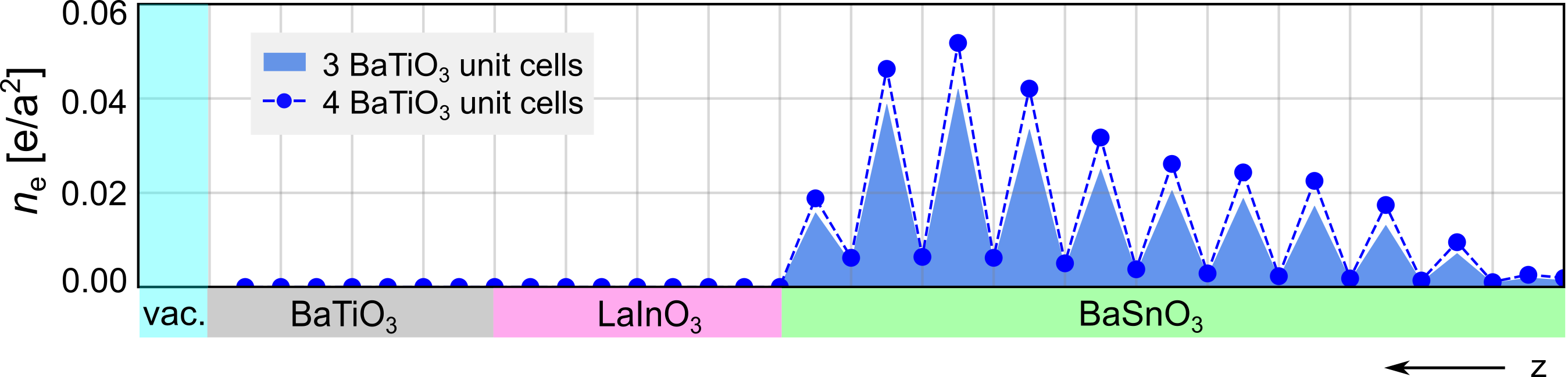}%
\renewcommand*{\thefigure}{S\arabic{figure}}
\caption{Distribution of electron charge within the BaSnO$_3$ side of the BTO/LIO/BSO heterostructure. The dashed line (blue area) refers to the case with 4 (3) BaTiO$_3$ unit cells.} 
\label{Fig-thickness}
 \end{center}
\end{figure*}

To understand how the thickness of the ferroelectric BaTiO$_3$ layer impacts the 2DEG density within the BaSnO$_3$ block, we compare the results for BTO/LIO/BSO heterostructures formed by 3 and 4 BaTiO$_3$ unit cells (u.c.). For the latter case, we choose a ferroelectric polarization of $P_{\downarrow}$=--0.04 C/m$^2$. Potential, \textit{V}, electric field, \textit{E}, and polarization, \textit{P}, are related as 
\begin{equation}
V=edE,\\
E=\frac{P}{\varepsilon_0 \varepsilon_r}, 
\end{equation}
where $\varepsilon_0$ and $\varepsilon_r$ are the dielectric permittivity of vacuum and the relative permittivity, respectively. To investigate the effect of thickness ($d$) independently, we ensure the same electric field induced within the BaTiO$_3$ block in both cases. Therefore, the following relation holds:
\begin{equation}
P_{3 u.c}=\frac{P_{4 u.c}d_{4 u.c}}{d_{3 u.c}}\sim -0.05~ C/m^{2}.\\
\end{equation}
We find that the 2DEG density increases with BaTiO$_3$ thickness from 0.26 $e/a^2$ (3 u.c.) to 0.33 $e/a^2$ (4 u.c.). The distribution of the 2DEG density within BaSnO$_3$ side exhibits charge accumulation with increasing thickness (Fig.~\ref{Fig-thickness}). Thus, in real samples where the thickness of the ferroelectric layer is tens of nanometers or more, we expect even larger 2DEGs densities under $P_{\downarrow}$ polarization, also considering that thicker layers allow for polarization  values above those chosen here.

\subsection{Heterostructure with relaxed $\mathrm{LaInO_{3}/BaSnO_{3}}$ interface }
\begin{figure*}
 \begin{center}
\includegraphics[width=.9\textwidth]{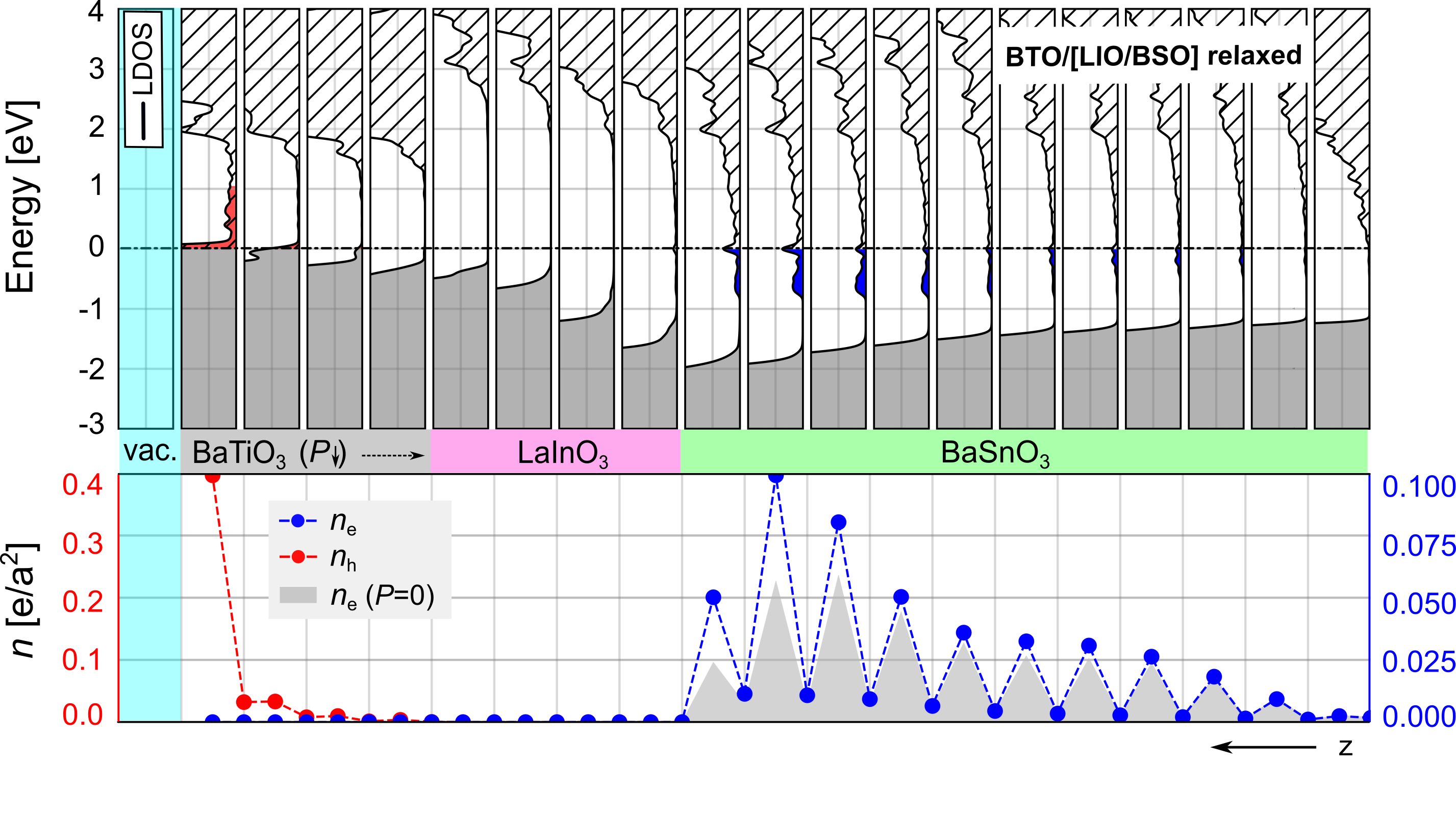}%
\renewcommand*{\thefigure}{S\arabic{figure}}
\caption{Electronic properties of the BTO/LIO/BSO heterostructure with relaxed LIO/BSO interface for a ferroelectric polarization of --0.12 C/m$^2$, \ie oriented towards the interface. The top panel shows the LDOS, where the gray area marks the occupied valence states (Fermi level set to zero). The depleted valence-band region (holes) and the occupied conduction-band region (electrons) resulting from electronic reconstruction are indicated by red and blue color, respectively. The bottom panel depicts the distribution of the electron (hole) charge density.}
\label{Fig-relax-pdown}
 \end{center}
\end{figure*}

\begin{figure*}
 \begin{center}
\includegraphics[width=.9\textwidth]{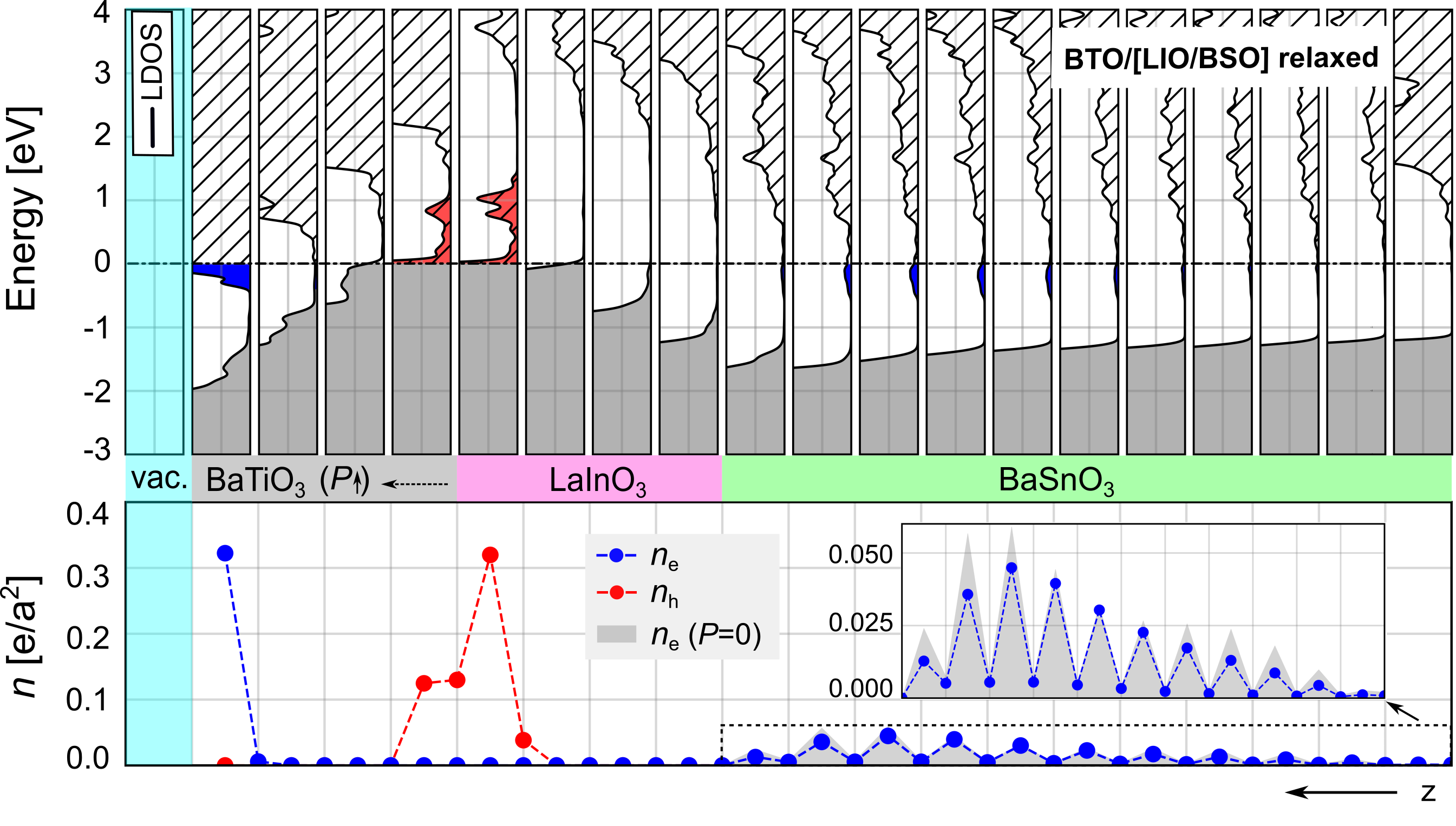}%
\renewcommand*{\thefigure}{S\arabic{figure}}
\caption{Same as Fig. \ref {Fig-relax-pdown} but with a ferroelectric polarization of +0.12 C/m$^2$ ($P_{\uparrow}$). For better visibility, the inset zooms into the 2DEG density within the BaSnO$_3$ side.}
\label{Fig-relax-pup}
 \end{center}
\end{figure*}

In Figs.~\ref{Fig-relax-pdown} and~\ref{Fig-relax-pup}, we depict the electronic properties of the relaxed LIO/BSO interface for both polarization directions, choosing their magnitude $\pm$0.12 C/m$^2$. The LDOS of the relaxed cases are very similar to their unrelaxed counterparts (see Figs.~4 and ~6 in the main text). Integrating the LDOS over the partially filled states within the BaSnO$_3$ block, we find a 2DEG density of about 0.49 $e/a^2$ (0.26 $e/a^2$) for the relaxed $P_{\downarrow}$ ($P_{\uparrow}$) case, compared to 0.37 $e/a^2$ for the relaxed paraelectric case. Overall, after relaxation, the trend of the 2DEG density as a function of polarization is more obvious (see Fig.~5 in the main text).

\end{document}